\documentclass[]{spie}  
\usepackage{float}
 
\usepackage{amsmath,amsfonts,amssymb}
\usepackage{graphicx}
\usepackage[colorlinks=true, allcolors=blue]{hyperref}

\title{Simulating Zernike-based CWFSing with HCIPy}

\author[a,b]{Klaus Subbotina Stephenson}
\author[b]{Kate Jackson}
\affil[a]{University of Victoria, 3800 Finnerty Rd Victoria, BC V8P 5C2, Canada}
\affil[b]{Herzberg Astronomy and Astrophysics Research Centre, 5071 W. Saanich Rd, Victoria, BC, V9E 2E7, Canada}

\authorinfo{Further author information: (Send correspondence to Klaus Subbotina Stephenson)\\Subbotina Stephenson, K: E-mail: klausss at uvic dot ca}

\begin{document} 
\maketitle

\begin{abstract}
This work explores optimal sensing distances in Zernike-based Curvature WaveFront Sensing (CWFS). By using \textit{HCIPy}, an open-source Python package, we simulate the first 100 Zernike modes at a variety of propagation distances ($1$ to $20,000$ m) to explore the relationship between the propagation of a wavefront and its spatial complexity to understand an idealized CWFS' sensitivity to individual modes. We find that there exist distinct optimal sensing distances for low-spatial-order-modes, with high-spatial-order modes demonstrating grouping behavior at exponentially shorter optimal sensing distances. Our results indicate that there is a sensitivity gap in current CWFSing that can be resolved by resorting to a high number of sensing planes for the low spatial order regime, with one sensing plane for each individual low-order mode; similarly, we propose several high-order regime sensing planes, with each sensing plane dedicated to small groups of localized modes centered around the highest modal sensitivity at a given propagation distance. We also discuss the caveats in dealing with high-order Zernike modes and the need for carrying out further propagation analysis in the Fourier space to extend preliminary work presented here. Lastly, we suggest the development of a new reconstruction algorithm that weighs the ability to reconstruct a complete wavefront according to the sensitivity (gain) to each mode at an individual sensing distance.
\end{abstract}

\keywords{Adaptive Optics, Wavefront Sensing, Curvature WaveFront Sensing (CWFS), Python, Fresnel propagation, Talbot effect, HCIPy, Simulations}

{\let\thefootnote\relax\footnotetext{© 2026 Society of Photo-Optical Instrumentation Engineers (SPIE): Astronomical Telescopes and Instrumentation. This version of "Simulating Zernike-based CWFSing with HCIPy" is made available here by the authors in agreement with the official SPIE web posting policy and is made public explicitly for noncommercial use. Klaus Subbotina Stephenson, Kate Jackson "Simulating Zernike-based CWFSing with HCIPy", Proc. SPIE 14150, Adaptive Optics Systems X, 141504Z (19 Aug 2026); https://doi.org/10.1117/12.3100745

}}

\section{INTRODUCTION}
\label{sec:intro}  
Adaptive Optics is a critical technology imperative for the success of ground-based High Contrast Imaging (HCI) astronomy. Exoplanet imaging, which requires extreme wavefront quality and stability compared to other sub-regimes of astronomy due to the orders of magnitude difference in the brightness of a host star and its planetary companion(s), has been significantly improved in near-IR to mid-IR wavelengths through the advent of modern AO and coronagraphic technologies \cite{Macintosh_Graham_Ingraham_Konopacky_Marois_Perrin_Poyneer_Bauman_Barman_Burrows_et, Jovanovic_2015}. While there have been several high-performance instruments that approach the fundamental diffraction limit, such as the Roman CoronaGraphic Imager, which reached $10^{-8}$ PSF contrasts in testing at JPL\cite{Roman_CGI}, VLT's SPHERE which achieves around $10^{-6}$ contrast\cite{SPHERE_VLT}, and MagAO-X which has achieved near $10^{-5}$ contrast \cite{2026arXiv260708146H}, a substantial portion of characterizing directly imaged exoplanets in recent years has been carried out by space-based telescopes as opposed to ground-based observatories \cite{JWST_performance, JWST_for_elt, TWA_20}. This is in part due to a gap in wavefront sensing stability and contrast currently afforded by ground-based instruments, which is the major motivation for this work. 

One hurdle in pushing wavefront sensing performance is grappling with the trade space between the variety of wavefront sensor used for a given AO system and that wavefront sensor's ability to sufficiently detect the range of spatial frequencies required for HCI science cases. For instance, the Shack-Hartmann Wavefront sensor, which is extremely robust and commonly used in many current adaptive optics systems \cite{Herriot_Morris_Anthony_Derdall_Duncan_Dunn_Ebbers_Fletcher_Hardy_Leckie_et, Neichel_2014, Wizinowich_Le_Mignant_Bouchez_Campbell_Chin_Contos_van_Dam_Hartman_Johansson_Lafon_et, Beuzit_paper}, is less sensitive to low spatial frequency aberrations due to its sub-aperture design\cite{Guyon_2005}. This prevents the detection of global tip/tilt terms, which are the largest atmospheric contributors per the Kolmogorov distribution. 
\section{Curvature WaveFront Sensing (CWFS)}
CWFS is a less-common form of wavefront sensing with inherently high sensitivity created by F. Roddier \cite{Roddier_Roddier_Roddier_1988} in 1988 and expanded upon significantly within the past two decades by multiple key works \cite{Guyon_2010, AOLI_nlCWFS_on_sky_results, Letchev_Crass_Crepp_Potier_2022}. CWFS, which takes place "out of" the pupil plane\footnote{Or focal plane, depending on the authors' preference. For this work, we will be operating "out of" the pupil plane.} uses two defocused pupil plane detections to measure the wavefront's curvature. The distances to each sensing plane are denoted as $L_+$ and $L_-$, and these intensity measurements on either side of the pupil plane are combined as follows to produce a single curvature measurement:
\begin{equation}
\label{curvature eq}
    C = \frac{L_+-L_-}{L_++L_-} 
\end{equation}
Where the difference between $L_+$ and $L_-$ is the focusing or defocusing of the pupil image, depending on which side of the pupil the measurement is taken. The found curvature constitutes the Laplacian of the original wavefront, which can then be used to solve Poisson's equation to estimate the original wavefront, or generally applied as measurements to a deformable mirror command matrix to actively correct turbulence in a closed-loop AO system.

Important to note for this work is the existence of "non-linear" CWFS (nlCWFS), which takes the CWFS concept further: as the $L_+$ and $L_-$ propagation distance is increased, the response of the system becomes non-linear. As proposed by O. Guyon in  2010 \cite{Guyon_2010}, a nlCWFS operates with the same setup as a CWFS but uses a reconstruction algorithm that both accounts for and utilizes the non-linear response space of CWFSing to achieve a higher dynamic range. As discussed later in section \ref{methods section}, the findings of this paper are applicable to both CWFS and nlCWFS.

\subsection{Fresnel propagation and the Talbot effect}

Due to its operation on either side of the pupil plane, CWFS relies on near-field Fresnel propagation to create defocus planes at which wavefront phase information is propagated far enough to be interpreted as intensity information. The optimal defocus distance at which a CWFS is most sensitive to a given turbulent mode, defined here using Zernike modes, is a critical consideration when building a CWFS. Finding optimal sensing distances in CWFS has been explored numerous times in the present CWFS literature, using a variety of methods such as simulating the number of iterations required to reduce the Root Mean Square (RMS) of an input phase error for a given sensing distance \cite{Letchev_Crass_Crepp_Potier_2022},  mathematically deriving theoretical distances using Fourier frequency analysis \cite{Huang_Xi_Liu_Jiang_2012}, and simulating various CWFS system parameters to determine a true-to-life methodology for finding optimal CWFS sensing distances \cite{Letchev_Crass_Crepp_2023}. Of these explorations into optimal sensing distances, a notable consensus is that the minimum sensing distance for CWFS relates to the Talbot effect for a given mode.

The Talbot length, a characteristic length at which the incoming wavefront is "re-imaged," is a near-field diffraction effect where monochromatic waves interfere in a cyclical pattern, resulting in phase-shifted self-imaging of the original wavefront. This effect occurs in both one-dimensional (coherent light through gratings) and two-dimensional (complex wavefronts) settings, and has been well documented in optical literature \cite{Talbot_effect_01, Talbot_effect_02}. The Talbot length is defined as follows:
\begin{equation}
    z_t=\frac{2d^2}{\lambda}
\end{equation}
Where the $z_t$ Talbot length is proportional to the square of \textit{d}, the pitch or spatial period of the input wavefront, divided by the operating wavelength, $\lambda$. For Zernike modes, we modify equation \ref{curvature eq} to become:
\begin{equation}
\label{zernike curvature approx}
    z_t=\frac{2(\frac{D}{n})^2}{\lambda}
\end{equation}
replacing $d$ with an approximation for Zernike spatial period. Here, $D$ is our pupil diameter and $n$ is the radial order of an individual Zernike mode. Equation \ref{zernike curvature approx} is used in this work as a rough approximation of the Talbot length within a Zernike context.
Previous work on CWFS sensing distances generally concludes the Talbot length should be used as a starting point to determine optimum sensing distances, as this cyclical re-imaging effect dictates where maximum contrast occurs for curvature signals in the propagation regimes closest to the pupil plane. However, for practical reasons, previous work deriving optimal CWFS distances relating to the direct calculation of the Talbot length has been limited, and there is a lack of a general methodology for calculating optimal sensing distances unspecific to any one CWFS system.


\section{Methods}
\label{methods section}
This paper takes a simple approach to finding optimal sensing distance for CWFS, one that does not involve a reconstruction algorithm, nor atmospheric turbulence simulations, and is thereby not limited in its application to CWFS nor nlCWFS alone: Single-step analysis of Zernike modes by comparing the system's ability to detect an individual mode in comparison to the original singularly input Zernike-mode phase aberration itself by comparing the recovered curvature signal RMS as a function of distance.

To produce these simulations and examine the relationship between spatial frequency and Fresnel propagated distance, we examine the first 100 Zernike modes using \textit{HCIPy}. \textit{HCIPy}, created by Emiel H. Por \cite{por2018hcipy}, is an open-source Python-based optical simulation package widely used for both coronagraphic and wavefront sensing scenarios. \textit{HCIPy} also provides a native Fresnel propagator, which is used here.
\subsection{Assumptions and considerations}
A major motivation for this work is to develop a simulation suite for CWFSing that will be translated into an optical design and eventually implemented on the 1.2-meter on-sky AO tested \textit{REVOLT}\cite{REVOLT}, located at NRC Herzberg in British Columbia, Canada. The largest customizations made to this Zernike-based propagation simulation are those changed to reflect \textit{REVOLT}'s operating parameters, i.e. the wavefront sensing wavelength used on-sky to close loops, the entrance pupil diameter, and system resolution. These values are $800$ nm, 1.2 m\footnote{We select to use the full telescope aperture for this work rather than scaling down to a test-bench space.}, and $352$ pixels, respectively, for the results provided here.
\subsection{Access}
\label{access section}
The simulations produced for this work will be made publicly available on the author's \hyperlink{https://github.com/caffeine-deprived/Simulating-Zernike-based-CWFSing-with-HCIPy}{GitHub}\footnote{Repository name: Simulating-Zernike-based-CWFSing-with-HCIPy}. The intention of sharing this work is to allow collaborators and interested readers to extend the research started here, and/or customize the code to desired parameters such that they may find optimal sensing distances for a particular CWFS system. 
\subsection{Simulations}
First making a Zernike basis using HCIPy's \textit{$make_-zernike_-basis$} function, we create a 100-mode-long basis using \textit{REVOLT}-specific parameters; the three aforementioned custom parameters can be easily swapped out to tailor the code to any CWFS system. These 100 modes are then looped through a list of various propagation distances, stretching from 1 meter to $2*10^6$ meters, to relate telescope space to the pupil plane\cite{Guyon_2008}. Each iteration of the loop then propagates a respective Zernike mode forward/backward away from the pupil plane using the \textit{$FresnelPropagator$} class. The intensity measurements on either side of the pupil plane are then combined, using equation \ref{curvature eq}, to produce a curvature measurement, which is then used to compute the curvature signal RMS in radians, effectively representing the power of the wavefront curvature at a given distance.

The rest of this subsection is split into three parts. The examination of explicit spatial complexity vs. propagation distance, found in section \ref{radial edge modes section}, covers radial-order edge modes with increasing spatial components while excluding azimuthal and symmetric modes. Section \ref{all detectable modes section} covers the remaining modes for completeness. Further detailed examination and plots for sub-populations of the first 100 Zernike modes in groups of increasing azimuthal complexity can be found in Appendix \ref{sec:bonus content}. Finally, section \ref{verification section} provides verification methodology for the simulations examined here.
\subsubsection{Radial-order edge modes}
\label{radial edge modes section}
\begin{figure}[H]
    \centering
    \includegraphics[width=0.9\linewidth]{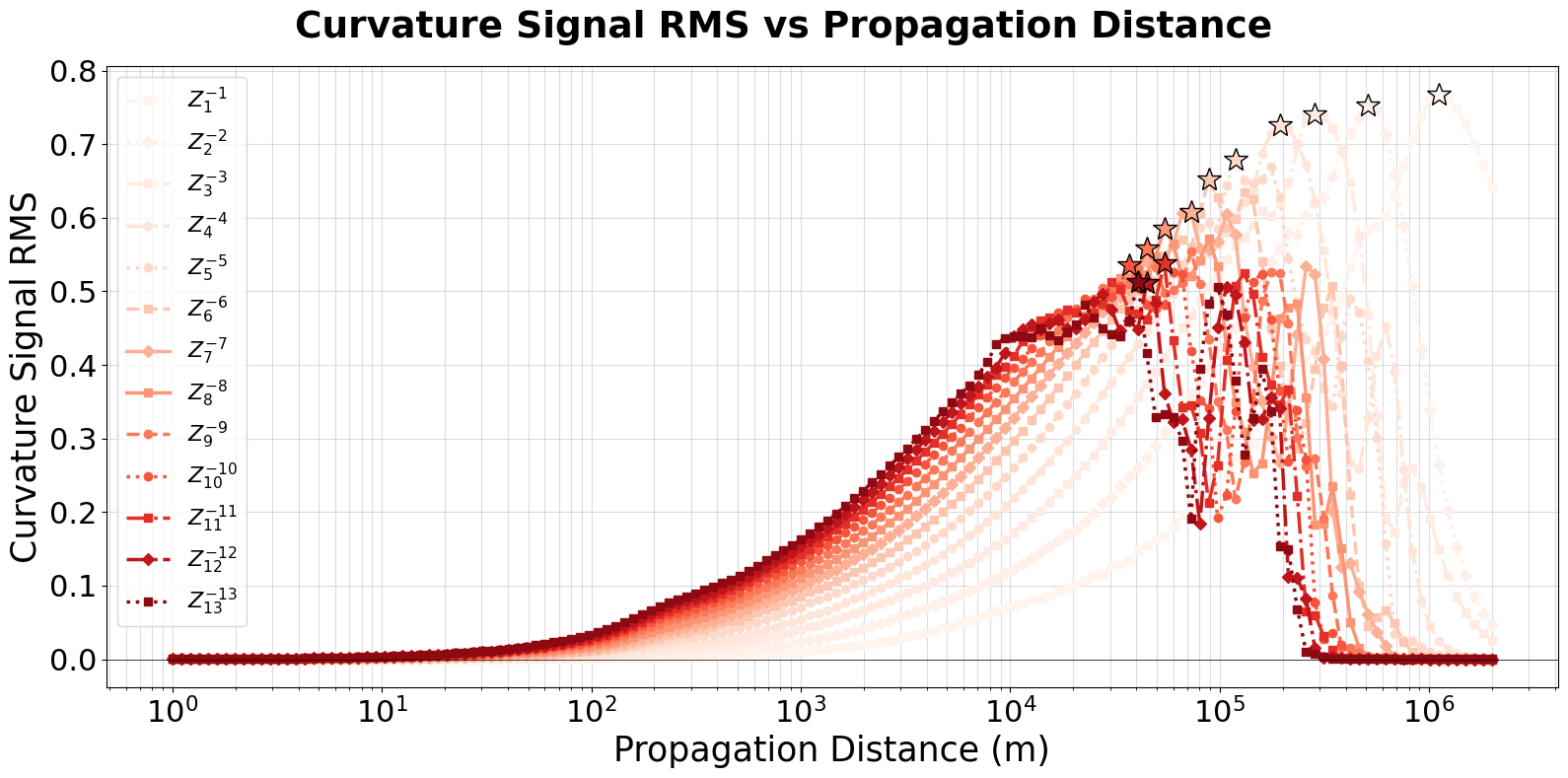}
    \caption{Plot displaying the propagation distance in meters versus the sensitivity to a specific mode in terms of curvature signal RMS. The radial-edge order modes plotted here correspond to 13 radial orders (excluding piston). The optimal sensing distance corresponds to where the sensitivity to a mode is equivalent to the highest obtainable curvature signal RMS. Additionally of note is the oscillating behavior visible for each of the 13 modes plotted here; this relates to the Talbot length for each mode and is visible in the plot as the oscillating curvature signal RMS over propagation distance.}
    \label{fig:radial_edge_curvature_v_propagation}
\end{figure}

\begin{figure}[H]
    \centering
    \includegraphics[width=0.9\linewidth]{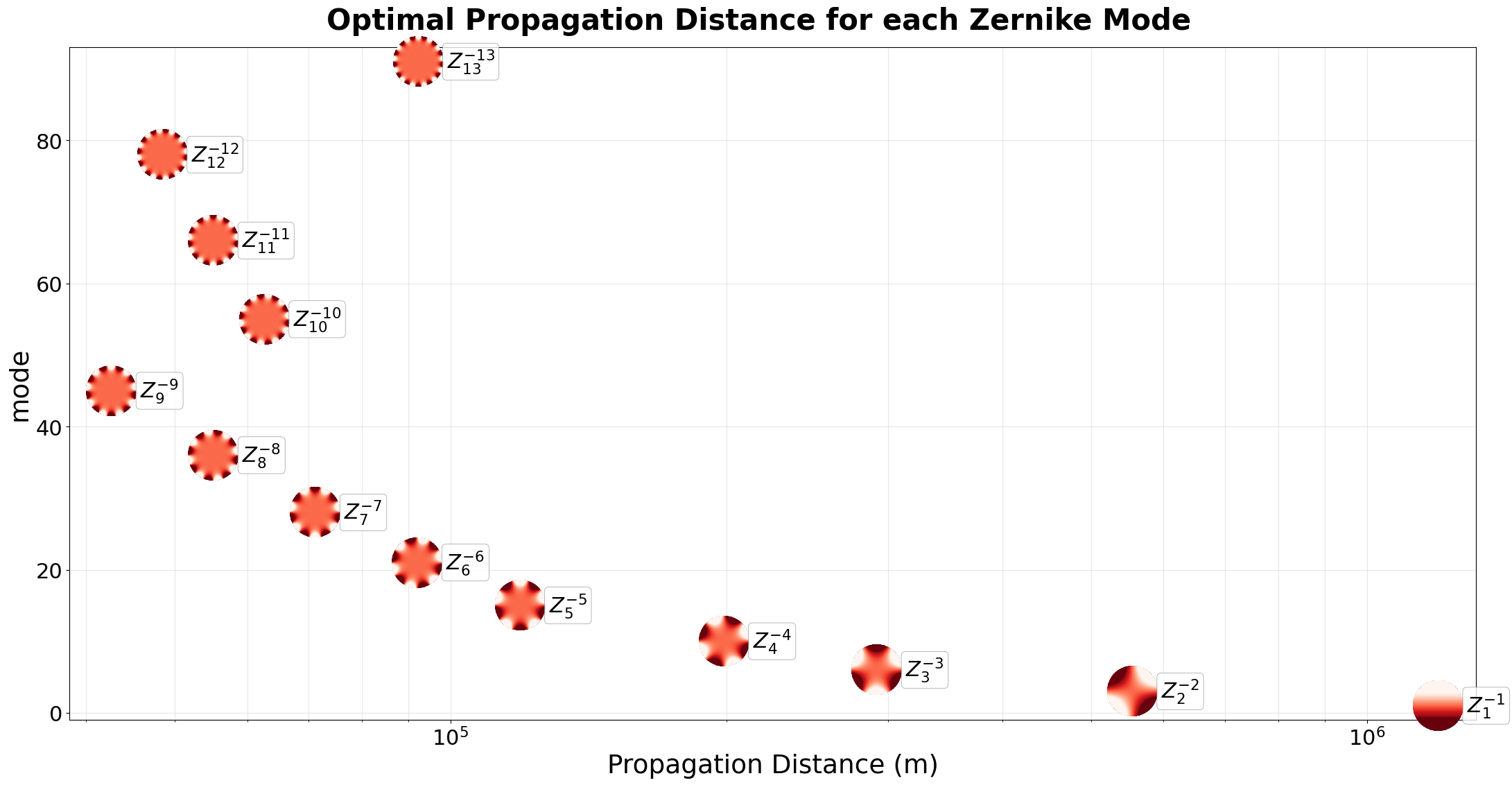}
    \caption{This figure shows the optimal propagation distance (in telescope space) for each radial-edge order mode up to the 13th radial order (excluding piston). The optimal propagation distance corresponds to the highest curvature signal RMS achieved in figure \ref{fig:radial_edge_curvature_v_propagation}.}
    \label{fig:radial_edge_optimal_distance_v_mode}
\end{figure}

Here we explicitly study the effects of increasing spatial complexity and the optimal propagation distance. By examining only radial-order edge modes, where $n=m$, we limit our results to studying modes whose radial and azimuthal order are equal. Figure \ref{fig:radial_edge_curvature_v_propagation} features the relationship between \textit{propagation distance} v. \textit{curvature signal RMS}, and Figure \ref{fig:radial_edge_optimal_distance_v_mode} shows the optimal propagation distance. The optimal propagation distance is equivalent to the location resulting in peak curvature signal RMS, as this denotes when phase is most efficiently transformed into intensity \cite{Transport_of_intensity}. 

\subsubsection{All modes}
\label{all detectable modes section}
In this section, we examine the first 100 Zernike modes in their entirety. Showcased in this section is the abundance of mixed spatial frequencies within the first 100 Zernike modes, which make it difficult to use the Talbot equation to determine an optimal propagation distance. This result and its implications are further expanded in section \ref{discussion section}. A color reference for each of the azimuthal orders is included in Appendix \ref{Zernike mode reference appendix} in Figure \ref{fig:zernike_rainbow}. 
\begin{figure}[H]
    \centering
    \includegraphics[width=0.9\linewidth]{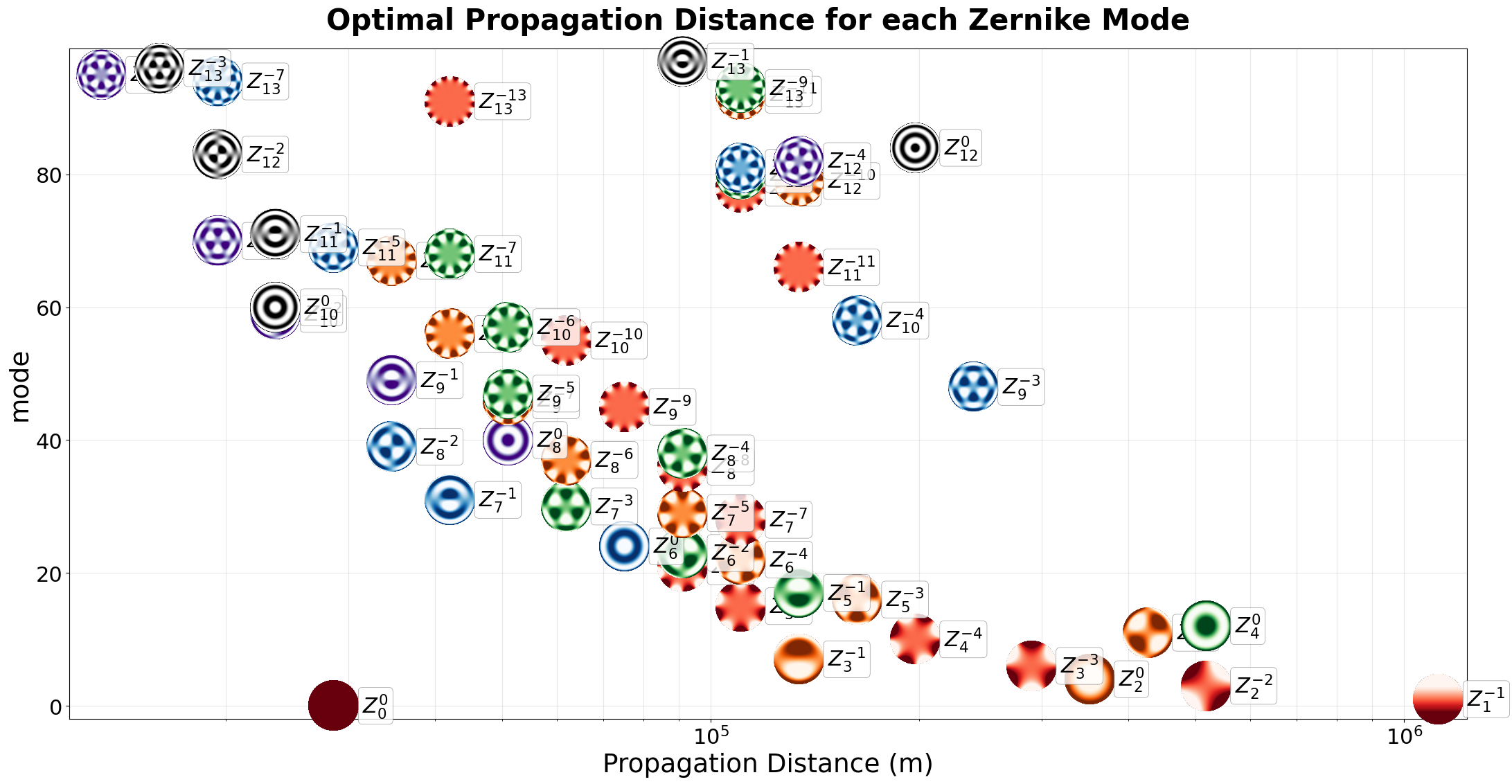}
    \caption{Optimal propagation distance versus mode, as colored by figure \ref{Zernike mode reference appendix} for the first 100 modes, including piston but excluding symmetric modes.}
    \label{fig:all_modes_optimal_d_v_mode}
\end{figure}

\subsubsection{Verification}
\label{verification section}
Creating an accurate Talbot length equation for the Zernike space, beyond the approximate equation \ref{zernike curvature approx}, is complex, and as such, we verify our found optimal sensing distances with a closed-loop verification. For this, we create another \textit{HCIPy}-based simulation which uses a modal (Zernike) DM that can intake observed curvature to update an input PSF. For this simulation, we use a DM with a 100-long (Zernike) modal basis, calibrated using a specific Fresnel propagation distance. The base interaction matrix is then inverted (pseudo-inverse using SVD decomposition), resulting in a command matrix which can then be updated with input wavefront curvature measurements to close the loop on any given wavefront.

For each of the modes examined in section \ref{all detectable modes section}, we apply the phase of a singular Zernike mode, with a coefficient of 1, to our input \textit{wavefront} object and propagate it\footnote{This propagation is idealized, with no atmospheric turbulence.} using the \textit{FresnelPropagator} to create intensity measurements. The intensity measurements are then used to compute curvature using equation \ref{curvature eq} and applied to our Zernike-based DM command matrix to close the loop on the original static singular input Zernike mode wavefront. The two dynamic parameters in this simulation are the calibration distance for the DM interaction matrix and the Fresnel propagation distance for the singular input phase wavefront.

\begin{figure}[H]
    \centering
    \includegraphics[width=0.9\linewidth]{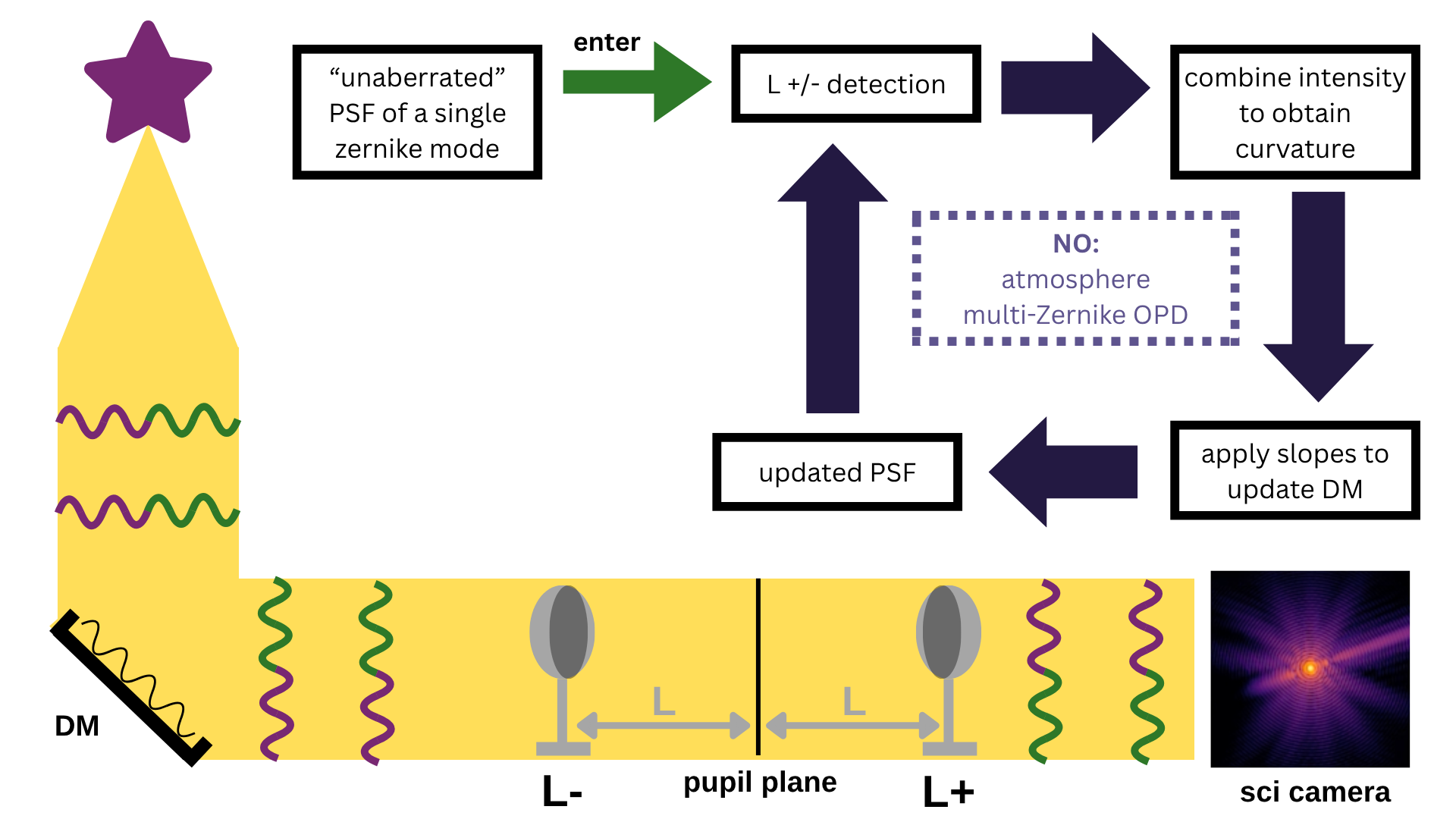}
    \caption{Diagram of the closed-loop simulation for verifying optimal propagation distances. Here, the initial wavefront to be corrected is the phase of an individual Zernike mode, which is then propagated towards/away from the pupil plane, converted into curvature using equation \ref{curvature eq}, and then applied to the distance-calibrated DM command matrix to close the loop on the original wavefront. This loop ends after 100 iterations, which is sufficient for an AO system correcting a static phase aberration. Depending on the calibration distance used to create the initial DM interaction matrix, we expect this closed-loop verification system to differ in performance as the calibration distance approaches the optimal sensing distance identified in section \ref{all detectable modes section}.}
    \label{fig:verification loop}
\end{figure}

\begin{figure}[H]
    \centering
    \includegraphics[width=0.9\linewidth]{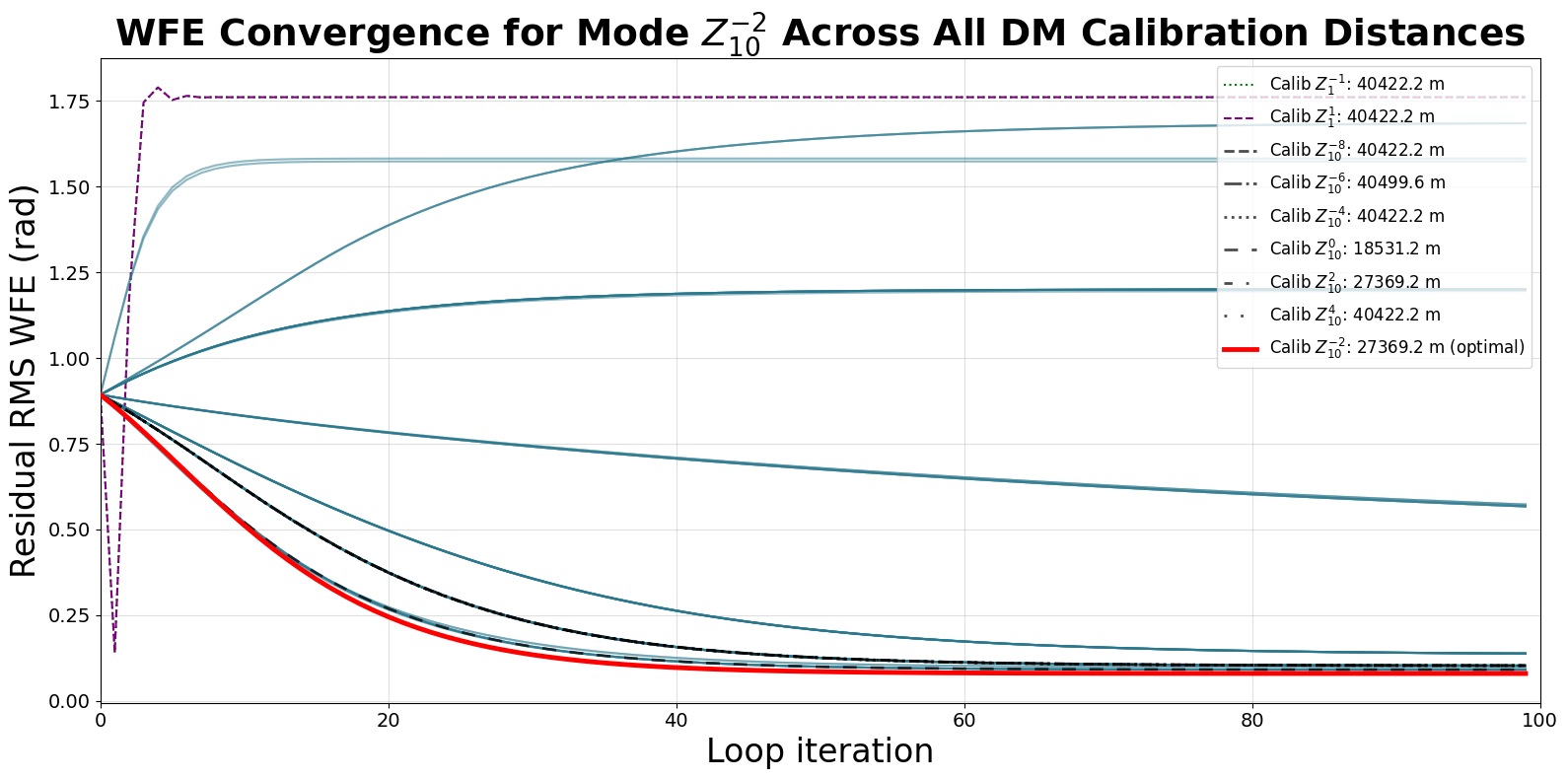}
    \caption{Example output of verifying the optimal sensing distance found in section \ref{methods section} for the Zernike mode $Z_{10}^{-2}$ (arbitrarily chosen). In this WaveFront Error (WFE) over time plot, we see the residual WFE (y-axis) decrease over 100 iterations (x-axis) of our unaberrated (lacking atmosphere and noise) closed-loop system. Each line represents a different DM calibration distance while all input phases are kept at a constant propagation distance; the constant propagation distance is the optimal sensing distance for the input mode found in section \ref{all detectable modes section}. The optimal calibration and propagation distance for mode $Z_{10}^{-2}$ is highlighted in red. The closest $\pm 3$ modes are labeled in black, with all other modal distances (of the 100 Zernike modes examined in this work) included in gray. Additionally labeled are the tip and tilt optimal sensing distances, as they starkly stand out due to their associated large sensing distances being insufficient for detecting mode $Z_{10}^{-2}$, a high-order mode with a significantly shorter optimal detection distance, as seen in Figure \ref{fig:purples}.}
    \label{fig:verification example}
\end{figure}

To verify that our found optimal distances for a given input Zernike phase are accurate, we run iterations of this closed-loop system with different DM calibration distances, looping through all 100 of the optimal distances found from simulation in section \ref{all detectable modes section}. For each calibrated DM interaction matrix, we close the loop on the same single input Zernike phase of interest, whose propagation distance within the closed loop remains constant across all iterations; this propagation distance is the optimal sensing distance for the Zernike mode being examined. We expect to see wavefront error reduce globally, or for the convergence time to decrease, the more accurate the DM calibration distance is relative to peak CWFS sensitivity for a given mode. A diagram of this verification process is found in Figure \ref{fig:verification loop}, and an example verification is included in Figure \ref{fig:verification example}.
\section{Discussion}
\label{discussion section}
Our results show that increasingly complex modes require shorter propagation distances. This agrees with current standards that complex modes require shorter defocus distances in CWFS, as over-propagating complex wavefronts causes the system to lose information per the transport-of-intensity equation \cite{Gureyev_Roberts_Nugent_1995}. Interestingly, this exponential relationship falls apart as azimuthal influence is included, as seen in Appendix \ref{Optimal Distances as a function of increasing azimuthal complexity appendix}. The simulations here additionally exhibit modal coupling, as seen in figures \ref{fig:all_modes_optimal_d_v_mode}, \ref{fig:oranges}, \ref{fig:greens} and \ref{fig:blues} where two modes of different complexities have the same optimal propagation distance (same x-axis location in the \textit{optimal propagation distance} v. \textit{mode} plot). As expected, the Talbot effect is present in each \textit{propagation distance} v. \textit{curvature Signal RMS} plot, the cyclical phase-shifted re-imaging of the original wavefront corresponding with oscillating changes in the measured curvature signal RMS. These results have two takeaways:
\begin{itemize}
    \item \textbf{Discrete low-order sensing distances:} for low-spatial order modes there appear to be distinct optimal sensing distances at which a CWFS or nlCWFS could be optimized for by placing a dedicated sensing plane at each low-order high-propagation distance. Further study is required to make a direct comparison as to how much sensitivity would be gained by CWFSing in having an individual sensing plane for each optimal low-order distance simulated here.
    \item \textbf{Modal Coupling:} Particularly present in the high-order regime, orders sharing the same or similar propagation distance (modes located at the same x-location in Figure \ref{fig:all_modes_optimal_d_v_mode}) prompt the ability to use a singular sensing plane for a small localized "group" of modes, as a single sensing distance here is optimal for multiple modes at once.
\end{itemize}
The validation architecture and example simulation for the found optimal distances, found in Figures \ref{fig:verification loop} and \ref{fig:verification example}, respectively, show improved WFE for the optimal sensing distance of a given Zernike mode. However, as seen in Figure \ref{fig:verification example}, there exist several other sensing distances that approach the found optimal sensing distance for mode $Z_{10}^{-2}$. This is likely due to both modal-crosstalk within the DM interaction matrix and the modal symmetries for this particular mode. 
With these three results in mind, \textbf{we suggest the creation of a high-planar CWFS/nlCWFS}. Having an infinite number of planes, one for each optimal sensing distance, would be the theoretical ideal CWFS/nlCWFS. By having a high number of measurement planes, one can optimize several of those planes to individual or small groups of low-order modes, pushing CWFS/nlCWFS closer to its theoretical sensitivity limit. For the high-order CWFS/nlCWFS regime, we suggest fewer planes of measurement and instead propose taking advantage of both coupling and the diminished gap in sensing distances, as one sensing plane located within this regime would be effectively sensitive to a higher number of modes as found in simulations here (compared to the lower- order mode regime). Additionally of note, a new reconstruction algorithm, beyond the standard Gerchberg-Saxton \cite{Gerchberg1972APA}, would be required for this theoretical best CWFS/nlCWFS, as a trade-off would be required to accommodate the volume of wavefront information received from a high-planar CWFS while still retaining the capability to operate as a time-resolved wavefront sensing method on-sky. While we do not explore alternative reconstruction algorithms here, we recognize that this proposed algorithm could use the simulations created here to create a weighted algorithm better optimized to the found sensing distances in this work. These results are preliminary; the authors will be utilizing both the lessons learned and the suite of simulations created here to expand further into creating a more advanced nlCWFSing simulation with \textit{HCIPy} and to inform the optical design of an upcoming nlCWFS for the on-sky \textit{REVOLT} testbed.

\section{Conclusion}
In this work, we have created an indirect method for finding optimal sensing distances for individual Zernike modes applicable to both CWFS and nlCWFS. This open-source (see section \ref{access section}) simulation depends on only three parameters, which can be easily swapped out to cater to any specific CWFS/nlCWFS system. Using our simulations, we found optimal sensing distances for the first 100 Zernike modes to examine the relationship between spatial complexity and propagation distance. The authors observe two key behaviors that can be used for optimizing CWFS/nlCWFS sensitivity: discrete low-order modal sensing distances and modal coupling. The former suggests a sensitivity space in CWFS/nlCWFS wherein individual sensing planes, one plane for each low-order mode, could improve the theoretical sensitivity limit of CWFSing. The latter finding indicates a unique ability for CWFSing to tailor its high-order sensing planes to localized groups of modes, capitalizing on both modal coupling and decreasing distances between optimal sensing locations. We thereby propose a new high-planar CWFS setup with several low-order sensing planes dedicated to individual modes and several high-order sensing planes intended for sensing small localized groups of high-order modes. 

Results show that Zernike modes, with their varying radial and azimuthal spatial frequency components, are poorly suited for the analysis conducted here (see Appendix \ref{Optimal Distances as a function of increasing azimuthal complexity appendix}). While this is currently understood to be due to modal-crosstalk and spatial aliasing, restricted in simulation by the resolution parameter in particular, the authors propose converting the simulations presented here to a Fourier basis in future work for more advanced analysis and verification. Lastly, we conclude that this work, while preliminary and requiring improvement, should be openly used and will be utilized by the authors in creating a high-planar nlCWFS optical design for the \textit{REVOLT} testbed in the coming years.

\appendix    

\section{Detailed assessments}
\label{sec:bonus content}
Here we include sub-groups of interest for the first 100 Zernike modes. Each group is organized by decreasing radial complexity and increasing azimuthal complexity, with section \ref{Optimal Distances as a function of increasing azimuthal complexity appendix} containing plots for subgroups of interest, and section \ref{Zernike mode reference appendix} containing a color-coded Zernike reference. Modes where $n+m \geq 8$ were excluded from this appendix as they provided no further insight into the conclusion that the wide range of mixed spatial frequencies exhibited by high-order Zernike modes makes finding optimal sensing distances difficult (and thereby requires a transition to Fourier space). 

\subsection{Optimal Distances as a function of increasing azimuthal complexity}
\label{Optimal Distances as a function of increasing azimuthal complexity appendix}
\begin{figure}[H]
    \centering
    \includegraphics[width=0.6\linewidth]{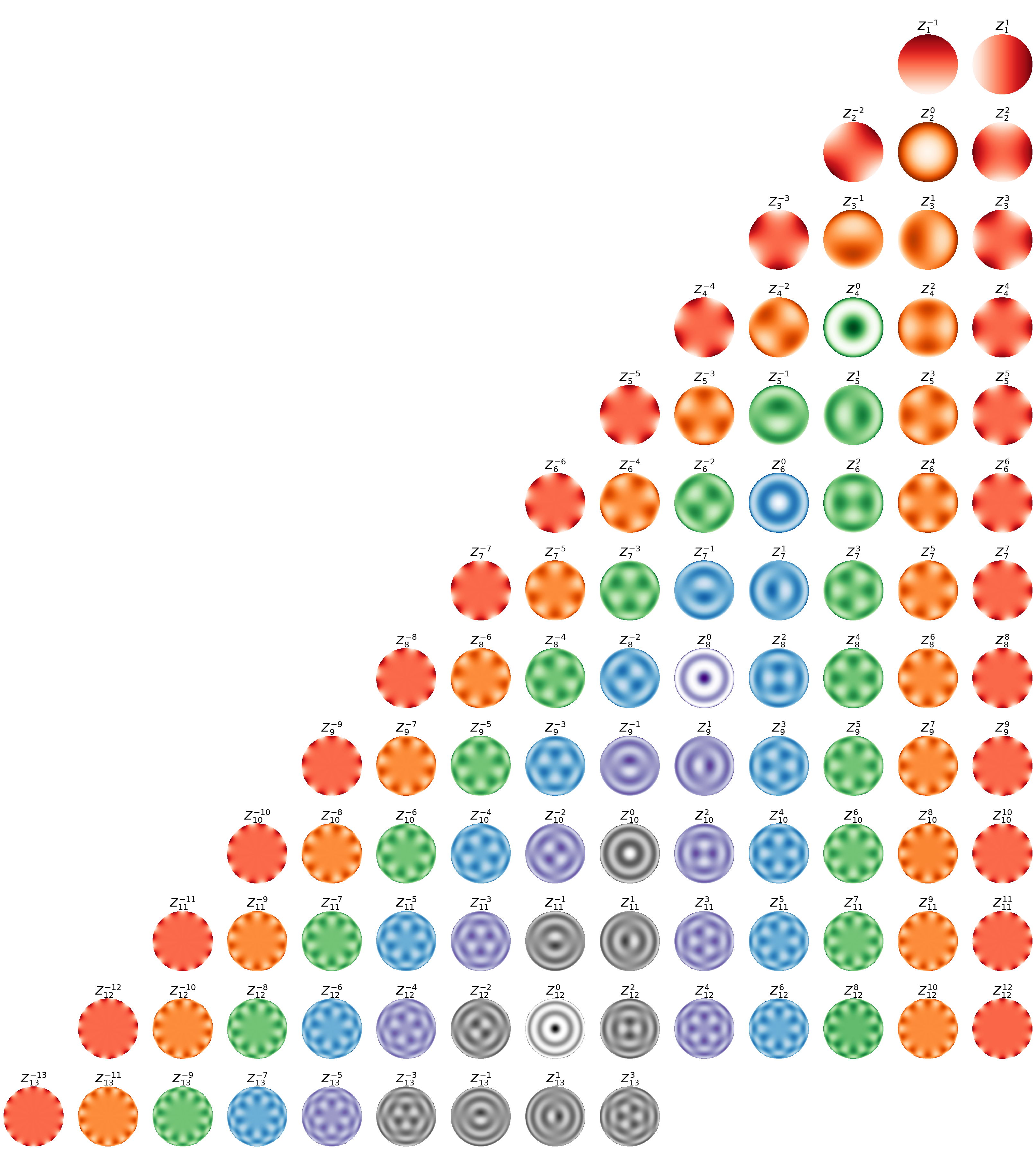}
    \caption{Colored reference for the first 100 Zernike modes as simulated in section \ref{all detectable modes section}, excluding piston.}
    \label{fig:zernike_rainbow}
\end{figure}

All subplots found in this section contain only the diagonal edge of Figure \ref{fig:zernike_rainbow} to reduce redundancy as was seen in figure \ref{fig:all_modes_optimal_d_v_mode} due to symmetric modes overlapping.

\begin{figure}[H]
    \centering
    \includegraphics[width=0.48\linewidth]{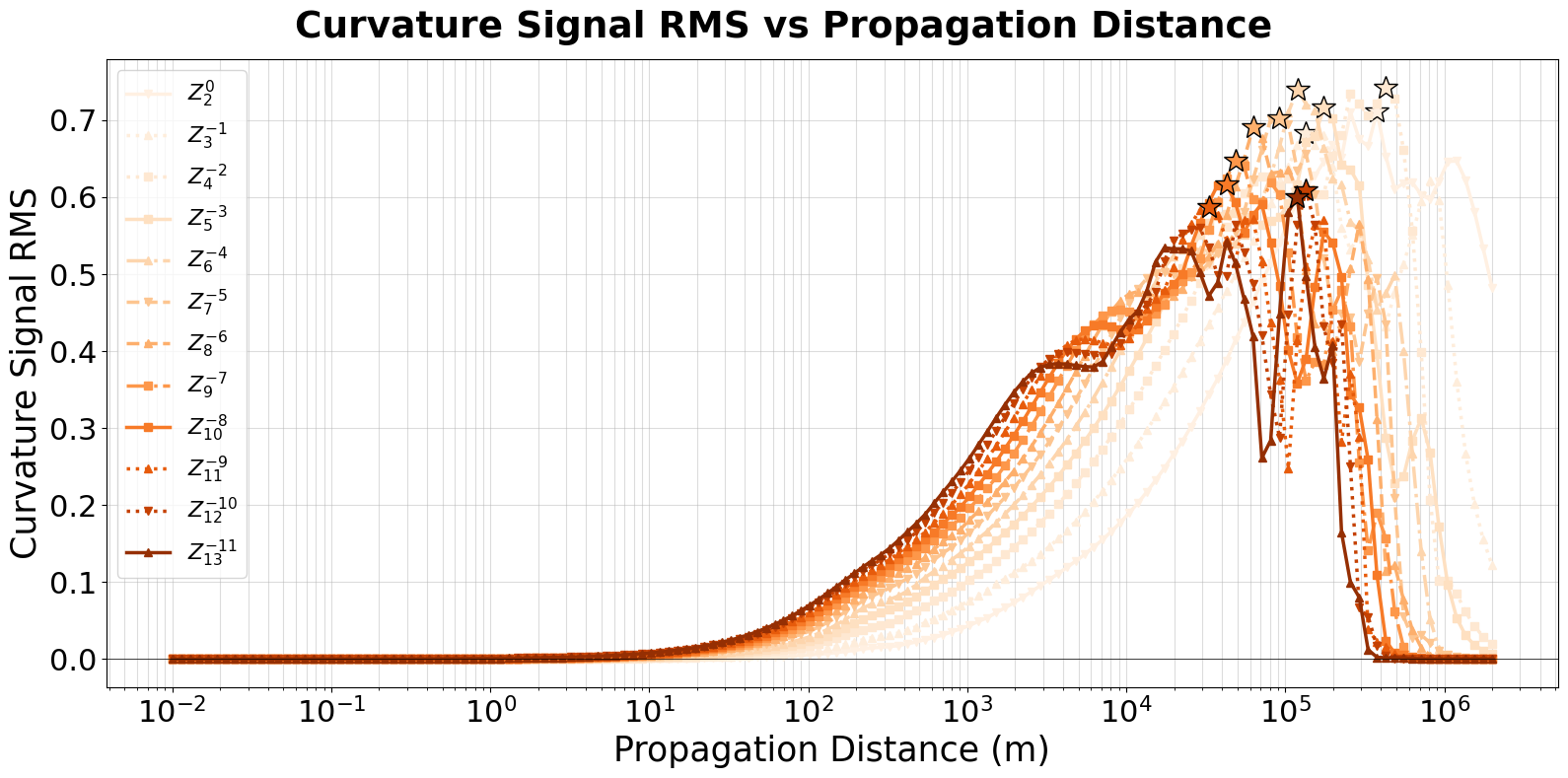}
    \includegraphics[width=0.48\linewidth]{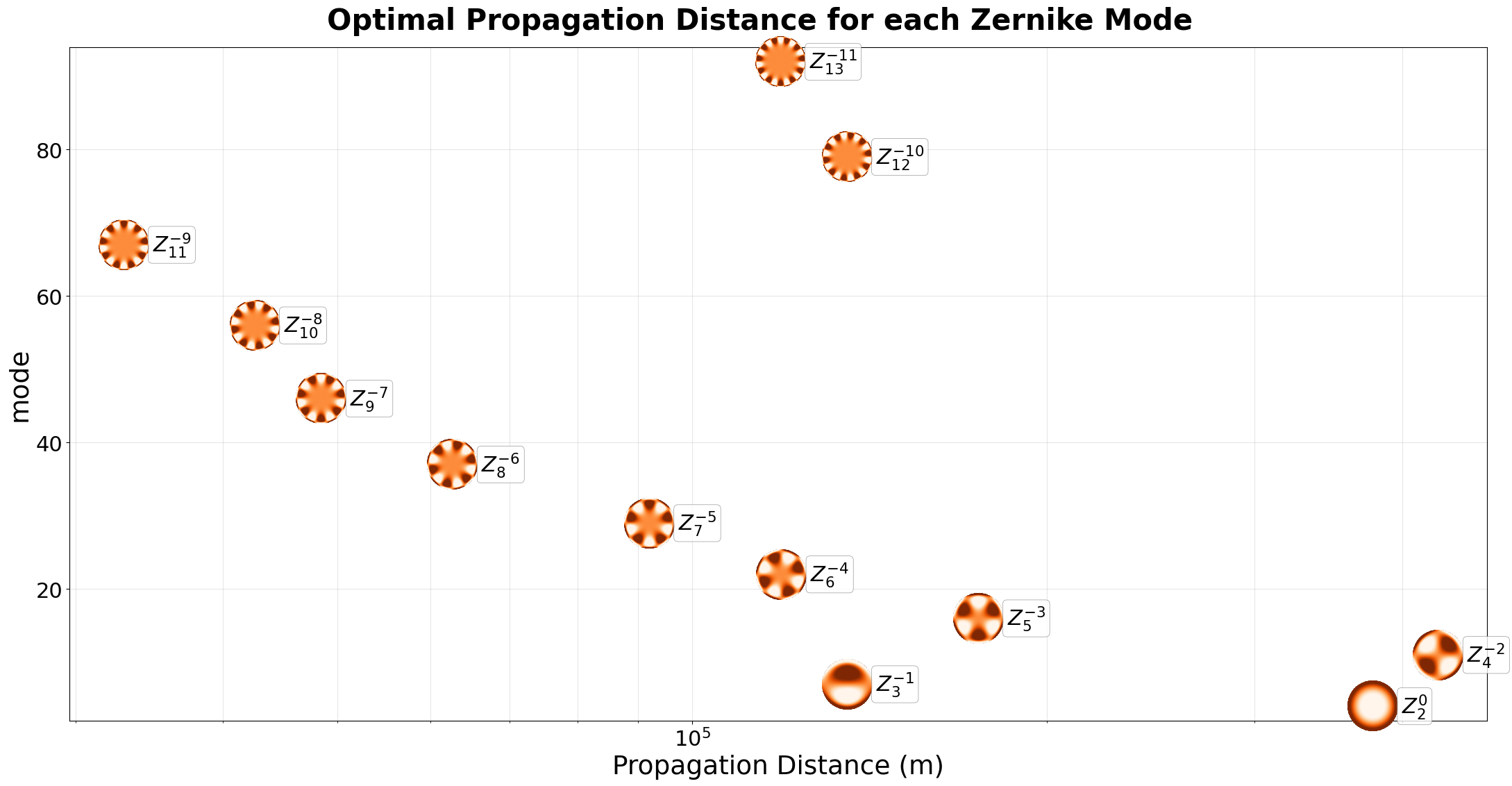}
    \caption{Comparison of propagation distance and curvature signal RMS for the second-lowest azimuthal order modes, excluding symmetric modes.}
    \label{fig:oranges}
\end{figure}

\begin{figure}[H]
    \centering
    \includegraphics[width=0.48\linewidth]{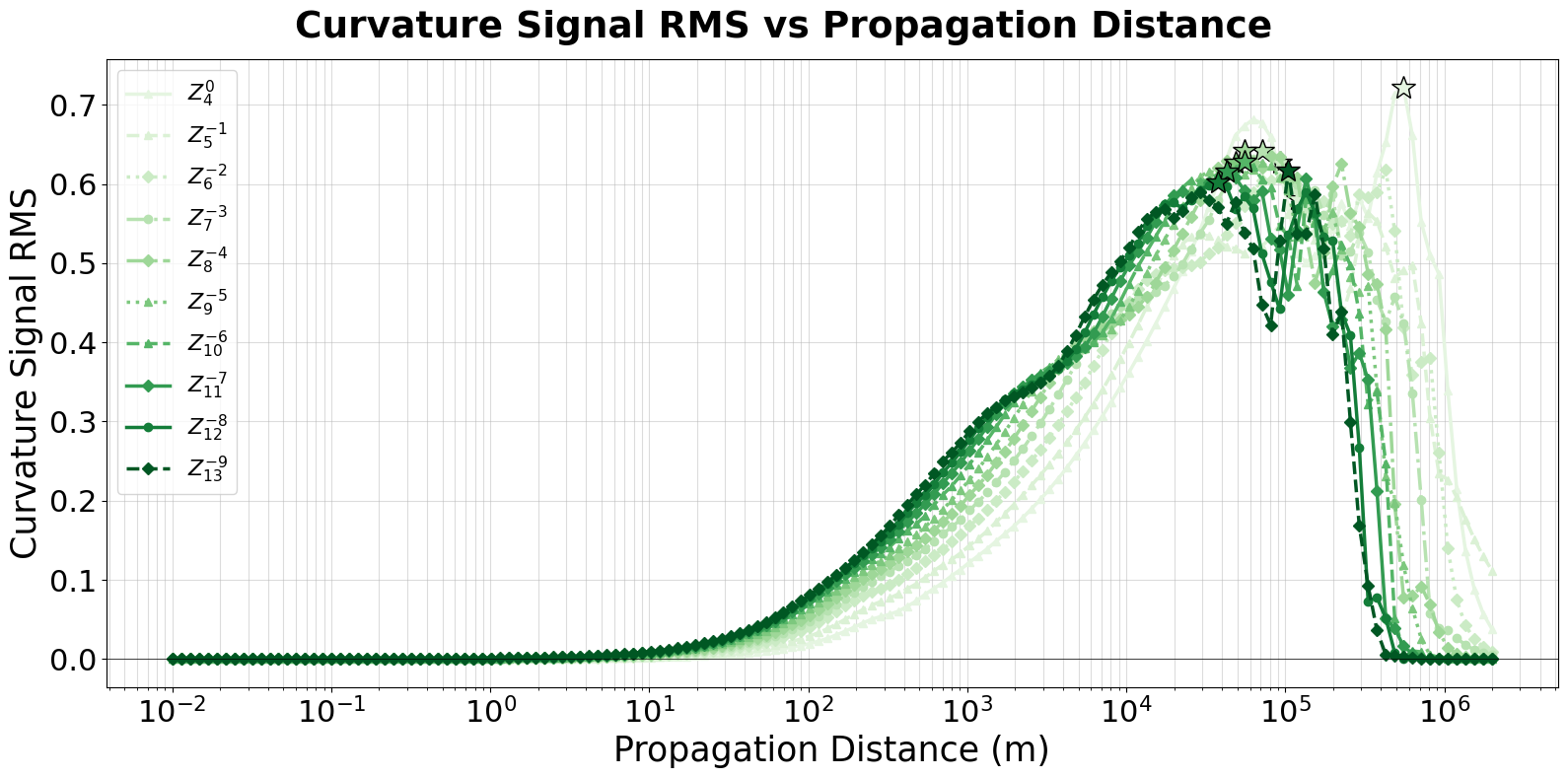}
    \includegraphics[width=0.48\linewidth]{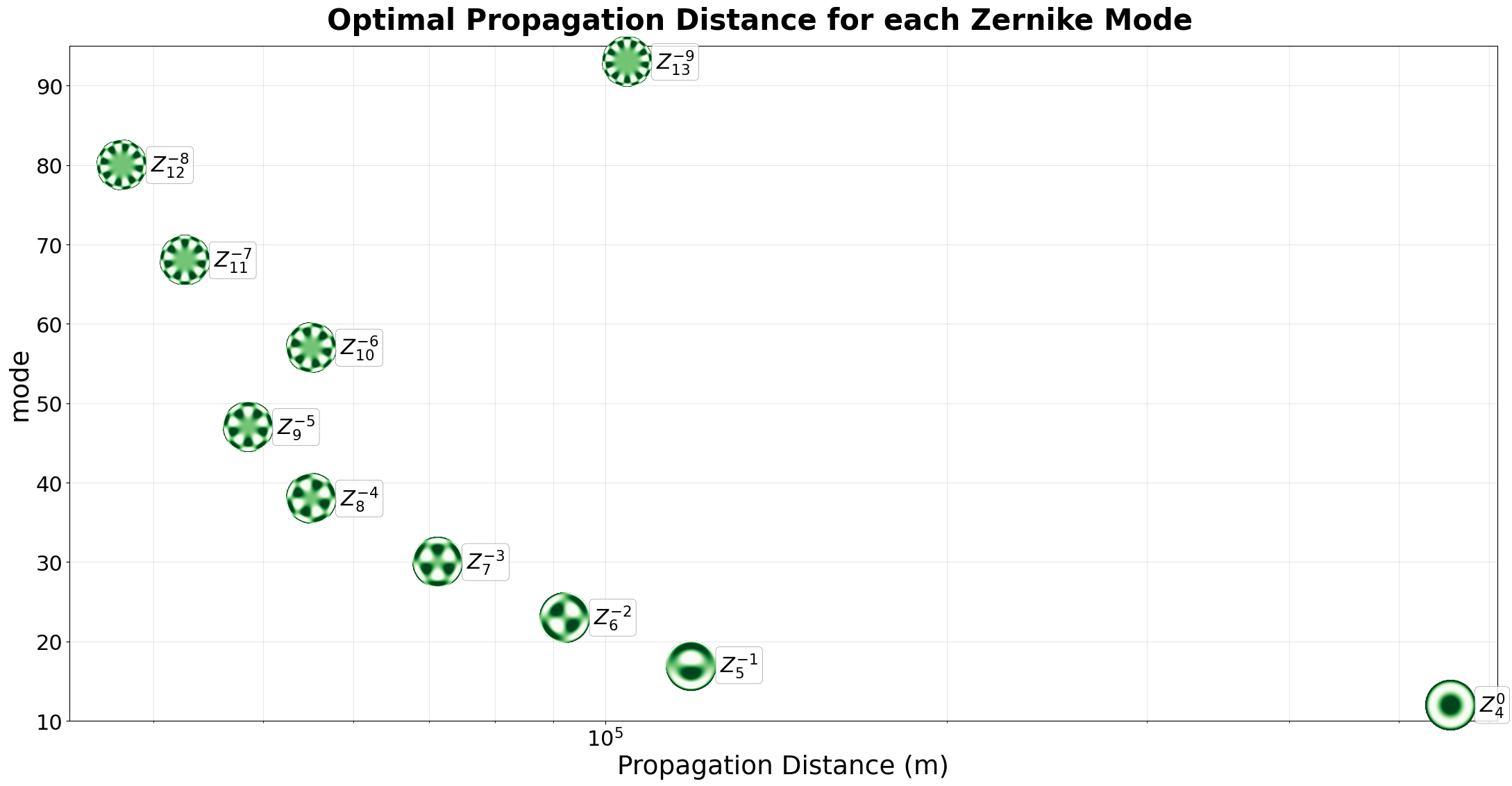}
    \caption{Comparison of propagation distance and curvature signal RMS for the third-lowest azimuthal order modes, excluding symmetric modes.}
    \label{fig:greens}
\end{figure}

\begin{figure}[H]
    \centering
    \includegraphics[width=0.48\linewidth]{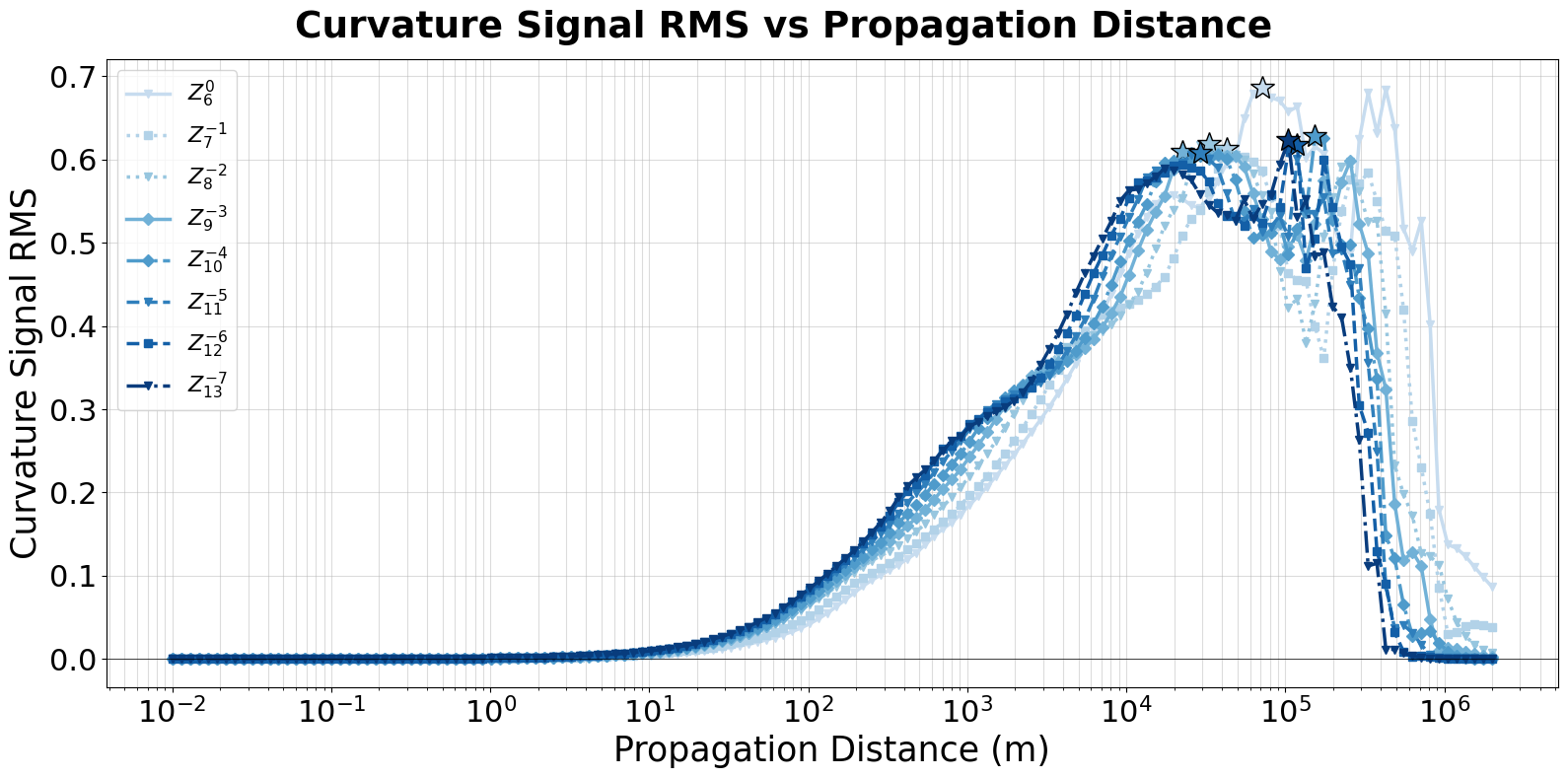}
    \includegraphics[width=0.48\linewidth]{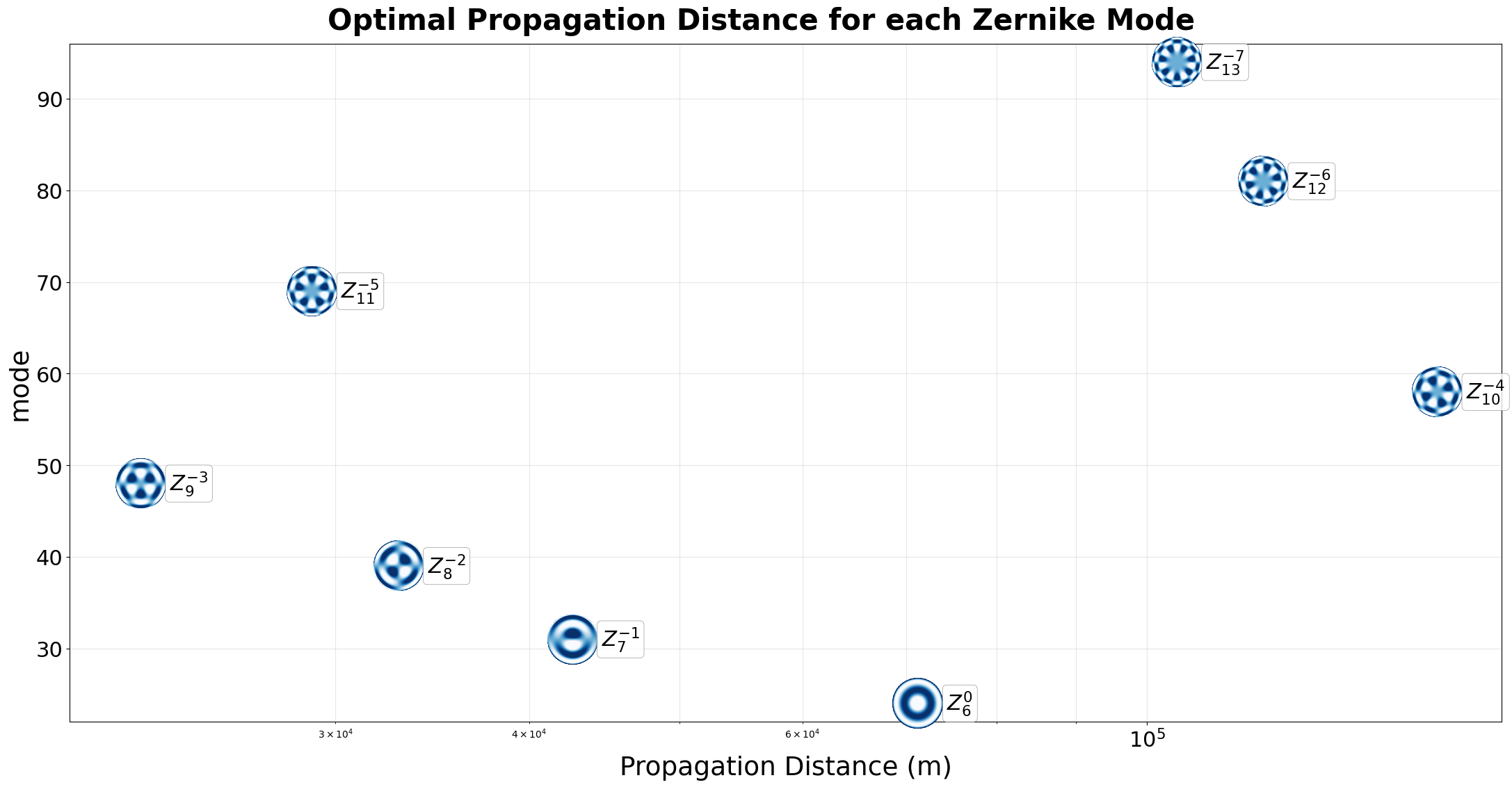}
    \caption{Comparison of propagation distance and curvature signal RMS for the second-highest azimuthal order modes included in this appendix, excluding symmetric modes.}
    \label{fig:blues}
\end{figure}

\begin{figure}[H]
    \centering
    \includegraphics[width=0.48\linewidth]{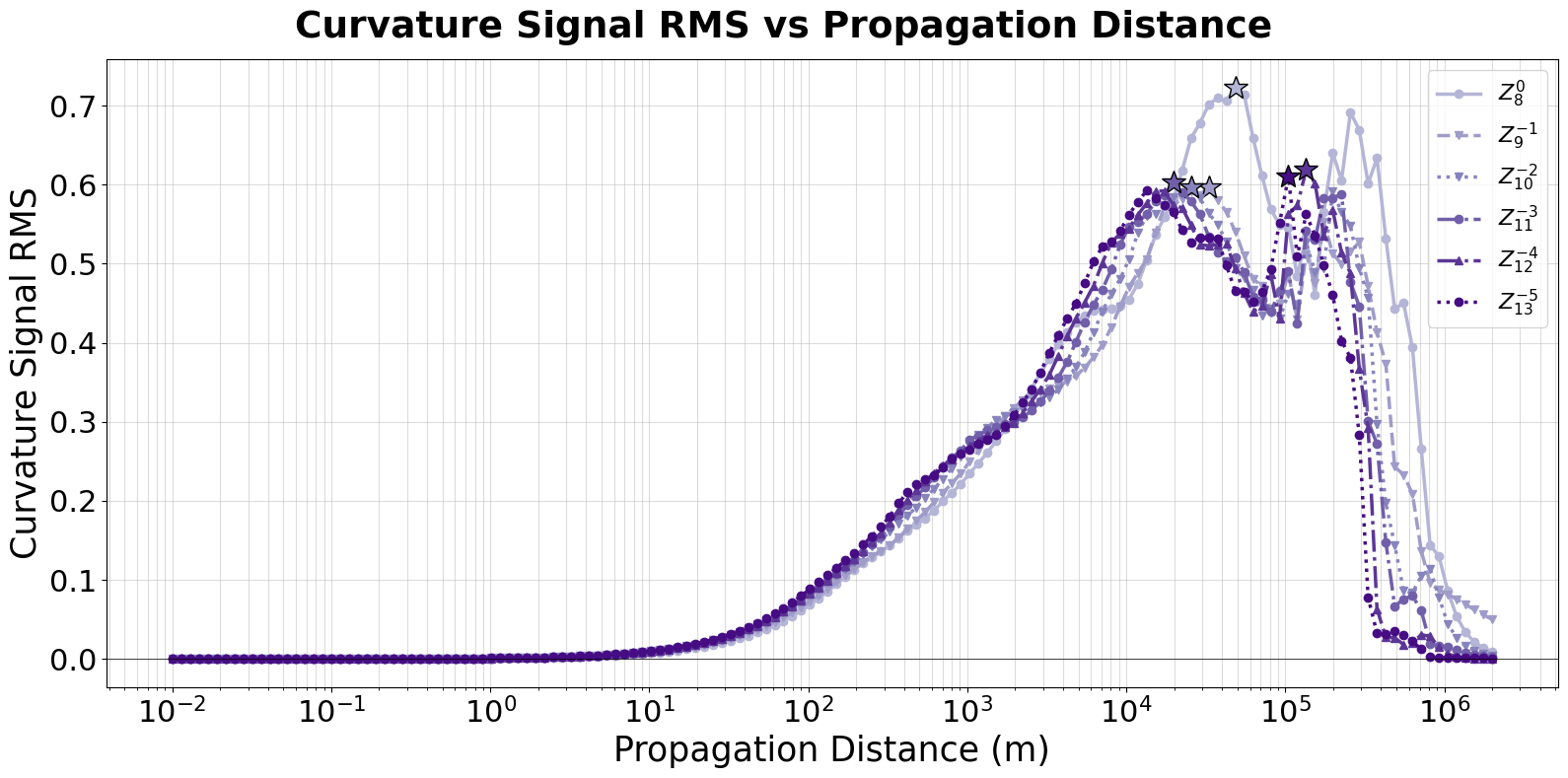}
    \includegraphics[width=0.48\linewidth]{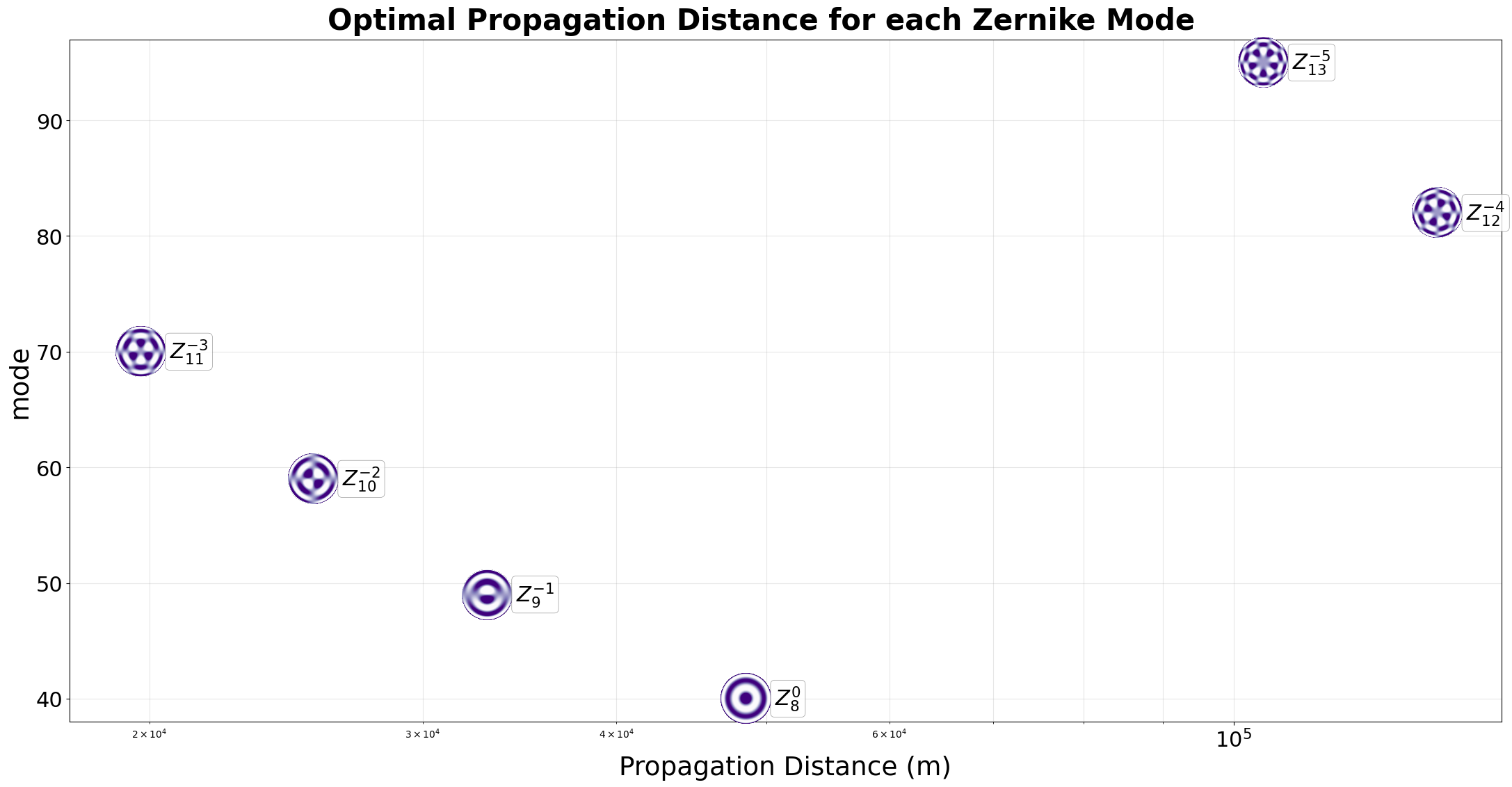}
    \caption{Comparison of propagation distance and curvature for the highest azimuthal order included in this appendix, excluding symmetric modes.}
    \label{fig:purples}
\end{figure}

\subsection{Zernike mode reference}
\label{Zernike mode reference appendix}
\acknowledgments 
The primary author would like to thank Maaike van Kooten, Peter Wizinowich, and Jean-Pierre Veran for their conversations and insight on various topics related to these proceedings.

\bibliography{report} 

@inproceedings{Roman_CGI,
    author  = {E., Cady}, 
    title   = "Milestone \#9 Report: Occulting Mask Coronagraph Dynamic Broadband Demonstration",
    publisher = "WFIRST Coronagraph Technology Development",
    year    = 2016
}

@article{SPHERE_VLT,
   title={Performance of the VLT Planet Finder SPHERE: II. Data analysis and results for IFS in laboratory},
   volume={576},
   ISSN={1432-0746},
   url={http://dx.doi.org/10.1051/0004-6361/201423910},
   DOI={10.1051/0004-6361/201423910},
   journal={Astronomy \&amp; Astrophysics},
   publisher={EDP Sciences},
   author={Mesa, D. and Gratton, R. and Zurlo, A. and Vigan, A. and Claudi, R. U. and Alberi, M. and Antichi, J. and Baruffolo, A. and Beuzit, J.-L. and Boccaletti, A. and Bonnefoy, M. and Costille, A. and Desidera, S. and Dohlen, K. and Fantinel, D. and Feldt, M. and Fusco, T. and Giro, E. and Henning, T. and Kasper, M. and Langlois, M. and Maire, A.-L. and Martinez, P. and Moeller-Nilsson, O. and Mouillet, D. and Moutou, C. and Pavlov, A. and Puget, P. and Salasnich, B. and Sauvage, J.-F. and Sissa, E. and Turatto, M. and Udry, S. and Vakili, F. and Waters, R. and Wildi, F.},
   year={2015},
   month=Apr, pages={A121} }

@misc{Talbot_effect_01,
      title={The Talbot Effect}, 
      author={Masud Mansuripur},
      year={2021},
      eprint={2103.06389},
      archivePrefix={arXiv},
      primaryClass={physics.optics},
      url={https://arxiv.org/abs/2103.06389}, 
}

@article{Talbot_effect_02, title={Bottlenecks of the wavefront sensor based on the Talbot effect}, volume={53}, DOI={10.1364/AO.53.00B223}, abstractNote={Physical constraints and peculiarities of the wavefront sensing technique, based on the Talbot effect, are discussed. The limitation on the curvature of the measurable wavefront is derived. The requirements to the Fourier spectrum of the periodic mask are formulated. Two kinds of masks are studied for their performance in the wavefront sensor. It is shown that the boundary part of the mask aperture does not contribute to the initial data for wavefront restoration. It is verified by experiment and computer simulation that the performance of the Talbot sensor, which meets established conditions, is similar to that of the Shack\&#x2013;Hartmann sensor.}, number={10}, journal={Appl. Opt.}, publisher={Optica Publishing Group}, author={Podanchuk, Dmytro and Kovalenko, Andrey and Kurashov, Vitalij and Kotov, Myhaylo and Goloborodko, Andrey and Danko, Volodymyr}, year={2014}, month=apr, pages={B223–B230} }

@article{Guyon_2008, title={Improving the Sensitivity of Astronomical Curvature Wavefront Sensor Using Dual-Stroke Curvature}, volume={120}, ISSN={1538-3873}, DOI={10.1086/589755}, abstractNote={Improving the Sensitivity of Astronomical Curvature Wavefront Sensor Using Dual-Stroke Curvature, Guyon, Olivier, Blain, Celia, Takami, Hideki, Hayano, Yutaka, Hattori, Masayuki, Watanabe, Makoto}, number={868}, journal={Publications of the Astronomical Society of the Pacific}, publisher={IOP Publishing}, author={Guyon, Olivier and Blain, Celia and Takami, Hideki and Hayano, Yutaka and Hattori, Masayuki and Watanabe, Makoto}, year={2008}, month=may, pages={655}, language={en} }

@article{Guyon_2010,
   title={High Sensitivity Wavefront Sensing with a Nonlinear Curvature Wavefront Sensor},
   volume={122},
   ISSN={1538-3873},
   url={http://dx.doi.org/10.1086/649646},
   DOI={10.1086/649646},
   number={887},
   journal={Publications of the Astronomical Society of the Pacific},
   publisher={IOP Publishing},
   author={Guyon, Olivier},
   year={2010},
   month=Jan, pages={49–62} }

@article{Huang_Xi_Liu_Jiang_2012, title={Frequency analysis of a wavefront curvature sensor: selection of propagation distance}, volume={59}, DOI={10.1080/09500340.2011.638741}, number={1}, journal={Journal of Modern Optics}, publisher={Taylor \& Francis}, author={Huang, Shengyang and Xi, Fengjie and Liu, Changhai and Jiang, Zongfu}, year={2012}, pages={35–41} }

@inproceedings{Roddier_Roddier_Roddier_1988, title={Curvature Sensing: A New Wavefront Sensing Method}, volume={0976}, url={https://www.spiedigitallibrary.org/conference-proceedings-of-spie/0976/0000/Curvature-Sensing-A-New-Wavefront-Sensing-Method/10.1117/12.948547.full}, DOI={10.1117/12.948547}, abstractNote={A new wavefront sensing method based on local wavefront curvature measurements rather than tilt measurements is investigated. Performances are compared to that of a Shack-Hartmann sensor. Examples of experimentally reconstructed wavefronts are presented and several applications are discussed.}, booktitle={Statistical Optics}, publisher={SPIE}, author={Roddier, Francois and Roddier, Claude and Roddier, Nicolas}, year={1988}, month=dec, pages={203–209} }

@inproceedings{AOLI_nlCWFS_on_sky_results, title={The AOLI low-order non-linear curvature wavefront sensor: laboratory and on-sky results}, volume={9148}, url={https://ui.adsabs.harvard.edu/abs/2014SPIE.9148E..2CC}, DOI={10.1117/12.2056137}, abstractNote={Many adaptive optics (AO) systems in use today require the use of bright reference objects to determine the effects of atmospheric distortions. Typically these systems use Shack-Hartmann Wavefront sensors (SHWFS) to distribute incoming light from a reference object between a large number of sub-apertures. Guyon et al. evaluated the sensitivity of several different wavefront sensing techniques and proposed the non-linear Curvature Wavefront Sensor (nlCWFS) offering improved sensitivity across a range of orders of distortion. On large ground-based telescopes this can provide nearly 100% sky coverage using natural guide stars. We present work being undertaken on the nlCWFS development for the Adaptive Optics Lucky Imager (AOLI) project. The wavefront sensor is being developed as part of a low-order adaptive optics system for use in a dedicated instrument providing an AO corrected beam to a Lucky Imaging based science detector. The nlCWFS provides a total of four reference images on two photon-counting EMCCDs for use in the wavefront reconstruction process. We present results from both laboratory work using a calibration system and the first on-sky data obtained with the nlCWFS at the 4.2 metre William Herschel Telescope, La Palma. In addition, we describe the updated optical design of the wavefront sensor, strategies for minimising intrinsic effects and methods to maximise sensitivity using photon-counting detectors. We discuss on-going work to develop the high speed reconstruction algorithm required for the nlCWFS technique. This includes strategies to implement the technique on graphics processing units (GPUs) and to minimise computing overheads to obtain a prior for a rapid convergence of the wavefront reconstruction. Finally we evaluate the sensitivity of the wavefront sensor based upon both data and low-photon count strategies.}, note={ADS Bibcode: 2014SPIE.9148E..2CC}, author={Crass, Jonathan and King, David and MacKay, Craig}, year={2014}, month=aug, pages={91482C} }

@inproceedings{Letchev_Crass_Crepp_Potier_2022, address={Montréal, Canada}, title={Spatial frequency response and sensitivity of the non-linear curvature wavefront sensor}, ISBN={978-1-5106-5351-1}, url={https://www.spiedigitallibrary.org/conference-proceedings-of-spie/12185/2632449/Spatial-frequency-response-and-sensitivity-of-the-non-linear-curvature/10.1117/12.2632449.full}, DOI={10.1117/12.2632449}, abstractNote={The nonlinear curvature wavefront sensor (nlCWFS) has been shown to be a promising alternative to existing wavefront sensor designs. Theoretical studies indicate that the inherent sensitivity of this device could oﬀer up to a factor of 10× improvement compared to the widely-used Shack-Hartmann wavefront sensor (SHWFS). The nominal nlCWFS design assumes the use of four detector measurement planes in a symmetric conﬁguration centered around an optical system pupil plane. However, the exact arrangement of these planes can potentially be optimized to improve aberration sensitivity, and minimize the number of iterations involved in the wavefront reconstruction process, and therefore reduce latency. We present a systematic exploration of the parameter space for optimizing the nlCWFS design. Using a suite of simulation tools, we study the eﬀects of measurement plane position on the performance of the nlCWFS and detector pixel sampling. A variety of seeing conditions are explored, assuming Kolmogorov turbulence. Results are presented in terms of residual wavefront error following reconstruction as well as the number of iterations required for solution convergence. Alternative designs to the symmetric four-plane design are studied, including three-plane and ﬁve-plane conﬁgurations. Finally, we perform a preliminary investigation of the eﬀects of broadband illumination on sensor performance relevant to astronomy and other applications.}, booktitle={Adaptive Optics Systems VIII}, publisher={SPIE}, author={Letchev, Stanimir and Crass, Jonathan and Crepp, Justin R. and Potier, Sam}, editor={Schmidt, Dirk and Schreiber, Laura and Vernet, Elise}, year={2022}, month=aug, pages={324}, language={en} }

@article{Letchev_Crass_Crepp_2023, title={Assessing Phase Reconstruction Accuracy for Different Nonlinear Curvature Wavefront Sensor Configurations}, volume={9}, ISSN={2329-4124}, url={http://arxiv.org/abs/2310.03062}, DOI={10.1117/1.JATIS.9.4.049001}, abstractNote={The nonlinear curvature wavefront sensor (nlCWFS) offers improved sensitivity for adaptive optics (AO) systems compared to existing wavefront sensors, such as the Shack-Hartmann. The nominal nlCWFS design uses a series of imaging planes offset from the pupil along the optical propagation axis as inputs to a numerically-iterative reconstruction algorithm. Research into the nlCWFS has assumed that the device uses four measurement planes configured symmetrically around the optical system pupil. This assumption is not strictly required. In this paper, we perform the first systematic exploration of the location, number, and spatial sampling of measurement planes for the nlCWFS. Our numerical simulations show that the original, symmetric four-plane configuration produces the most consistently accurate results in the shortest time over a broad range of seeing conditions. We find that the inner measurement planes should be situated past the Talbot distance corresponding to a spatial period of $r_0$. The outer planes should be large enough to fully capture field intensity and be situated beyond a distance corresponding to a Fresnel-number-scaled equivalent of $Zapprox50$ km for a $D=0.5$ m pupil with $λ=532$ nm. The minimum spatial sampling required for diffraction-limited performance is 4-5 pixels per $r_0$ as defined in the pupil plane. We find that neither three-plane nor five-plane configurations offer significant improvements compared to the original design. These results can impact future implementations of the nlCWFS by informing sensor design.}, note={arXiv:2310.03062 [astro-ph]}, number={04}, journal={Journal of Astronomical Telescopes, Instruments, and Systems}, author={Letchev, Stanimir and Crass, Jonathan and Crepp, Justin R.}, year={2023}, month=oct }

@inproceedings{por2018hcipy,
    author = {Por, E.~H. and Haffert, S.~Y. and Radhakrishnan, V.~M. and Doelman, D.~S. and Van Kooten, M. and Bos, S.~P.},
    title = "{High Contrast Imaging for Python (HCIPy): an open-source adaptive optics and coronagraph simulator}",
    booktitle = {Adaptive Optics Systems VI},
    year = 2018,
    series = {Proc. {{SPIE}}},
    volume = 10703,
    doi = {10.1117/12.2314407},
    URL = {https://doi.org/10.1117/12.2314407}
}

@ARTICLE{Transport_of_intensity,
       author = {{Zuo}, Chao and {Li}, Jiaji and {Sun}, Jiasong and {Fan}, Yao and {Zhang}, Jialin and {Lu}, Linpeng and {Zhang}, Runnan and {Wang}, Bowen and {Huang}, Lei and {Chen}, Qian},
        title = "{Transport of intensity equation: a tutorial}",
      journal = {Optics and Lasers in Engineering},
         year = 2020,
        month = dec,
       volume = {135},
          eid = {106187},
        pages = {106187},
          doi = {10.1016/j.optlaseng.2020.106187},
       adsurl = {https://ui.adsabs.harvard.edu/abs/2020OptLE.13506187Z}
}

@article{Gerchberg1972APA,
  title={A practical algorithm for the determination of phase from image and diffraction plane pictures},
  author={R. W. Gerchberg},
  journal={Optik},
  year={1972},
  volume={35},
  pages={237-246},
  url={https://api.semanticscholar.org/CorpusID:55691159}
}

@article{JWST_performance,
   title={The Science Performance of JWST as Characterized in Commissioning},
   volume={135},
   ISSN={1538-3873},
   url={http://dx.doi.org/10.1088/1538-3873/acb293},
   DOI={10.1088/1538-3873/acb293},
   number={1046},
   journal={Publications of the Astronomical Society of the Pacific},
   publisher={IOP Publishing},
   author={Rigby, Jane et al},
   year={2023},
   month=Apr, pages={048001} }

@INPROCEEDINGS{JWST_for_elt,
       author = {{Biller}, Beth and {Carter}, A. and {Ferrer-Chavez}, R. and {Kane}, R. and {Bogat}, E. and {Strampelli}, G. and {Girard}, J. and {Lawson}, K. and {Wang}, J. and {Booth}, M. and {Vos}, J. and {Zhou}, Y. and {Fontanive}, C.},
        title = "{JWST direct imaging searches for sub-Jupiter planets as a pathfinder towards ELT exoplanet imaging}",
    booktitle = {Planetary formation and Exoplanets in the ELT era (Exo-ELT)},
         year = 2026,
        month = apr,
          eid = {54},
        pages = {54},
          doi = {10.5281/zenodo.19686485},
       adsurl = {https://ui.adsabs.harvard.edu/abs/2026exoe.confE..54B}
}

@ARTICLE{TWA_20,
       author = {{Palatnick}, Skyler and {Millar-Blanchaer}, Maxwell A. and {Zhang}, Jingwen and {Lawson}, Kellen and {Lewis}, Briley L. and {Crotts}, Katie A. and {Carter}, Aarynn L. and {Biller}, Beth and {Girard}, Julien H. and {Marino}, Sebastian and {Bendahan-West}, Rapha{\"e}l and {Strampelli}, Giovanni M. and {James}, Andrew D. and {Stephenson}, Klaus Subbotina and {Ferrer-Chavez}, Rodrigo and {Booth}, Mark and {Sutlieff}, Ben J. and {Sanghi}, Aniket and {Fontanive}, Cl{\'e}mence and {Rickman}, Emily and {Rebollido}, Isabel and {Hoch}, Kielan and {Balmer}, William O.},
        title = "{Discovery of a Debris Disk around TWA 20}",
      journal = {\apj},
         year = 2025,
        month = dec,
       volume = {995},
       number = {2},
          eid = {149},
        pages = {149},
          doi = {10.3847/1538-4357/ae1963},
archivePrefix = {arXiv},
       eprint = {2510.20216},
 primaryClass = {astro-ph.EP},
       adsurl = {https://ui.adsabs.harvard.edu/abs/2025ApJ...995..149P}
}

@ARTICLE{Guyon_2005,
       author = {{Guyon}, Olivier},
        title = "{Limits of Adaptive Optics for High-Contrast Imaging}",
      journal = {\apj},
         year = 2005,
        month = aug,
       volume = {629},
       number = {1},
        pages = {592-614},
          doi = {10.1086/431209},
archivePrefix = {arXiv},
       eprint = {astro-ph/0505086},
 primaryClass = {astro-ph},
       adsurl = {https://ui.adsabs.harvard.edu/abs/2005ApJ...629..592G}
}

@article{Gureyev_Roberts_Nugent_1995, title={Partially coherent fields, the transport-of-intensity equation, and phase uniqueness}, volume={12}, DOI={10.1364/JOSAA.12.001942}, abstractNote={Recent papers have shown that there are different coherent and partially coherent fields that may have identical intensity distributions throughout space. On the other hand, the well-known transport-of-intensity equation allows the phase of a coherent field to be recovered from intensity measurements, and the solution is widely held to be unique. A discussion is given on the recovery of the structure of both coherent and partially coherent fields from intensity measurements, and we reconcile the uniqueness question by showing that the transport-of-intensity equation has a unique solution for the phase only if the intensity distribution has no zeros.}, number={9}, journal={J. Opt. Soc. Am. A}, publisher={Optica Publishing Group}, author={Gureyev, T. E. and Roberts, A. and Nugent, K. A.}, year={1995}, pages={1942–1946} }

@article{Macintosh_Graham_Ingraham_Konopacky_Marois_Perrin_Poyneer_Bauman_Barman_Burrows_et, title={First light of the Gemini Planet Imager}, volume={111}, DOI={10.1073/pnas.1304215111}, abstractNote={Significance
Direct detection—spatially resolving the light of a planet from the light of its parent star—is an important technique for characterizing exoplanets. It allows observations of giant exoplanets in locations like those in our solar system, inaccessible by other methods. The Gemini Planet Imager (GPI) is a new instrument for the Gemini South telescope. Designed and optimized only for high-contrast imaging, it incorporates advanced adaptive optics, diffraction control, a near-infrared spectrograph, and an imaging polarimeter. During first-light scientific observations in November 2013, GPI achieved contrast performance that is an order of magnitude better than conventional adaptive optics imagers.}, journal={Proceedings of the National Academy of Sciences}, author={Macintosh, Bruce and Graham, James and Ingraham, Patrick and Konopacky, Quinn and Marois, Christian and Perrin, Marshall and Poyneer, Lisa and Bauman, Brian and Barman, Travis and Burrows, Adam and Cardwell, Andrew and Chilcote, Jeffrey and De Rosa, Robert and Dillon, Daren and Doyon, Rene and Dunn, Jennifer and Erikson, Darren and Fitzgerald, Michael and Gavel, Donald and Wolff, Schuyler}, year={2014}, month=may, pages={12661–12666} }

@article{Jovanovic_2015,
   title={The Subaru Coronagraphic Extreme Adaptive Optics System: Enabling High-Contrast Imaging on Solar-System Scales},
   volume={127},
   ISSN={1538-3873},
   url={http://dx.doi.org/10.1086/682989},
   DOI={10.1086/682989},
   number={955},
   journal={Publications of the Astronomical Society of the Pacific},
   publisher={IOP Publishing},
   author={Jovanovic, N. and Martinache, F. and Guyon, O. and Clergeon, C. and Singh, G. and Kudo, T. and Garrel, V. and Newman, K. and Doughty, D. and Lozi, J. and Males, J. and Minowa, Y. and Hayano, Y. and Takato, N. and Morino, J. and Kuhn, J. and Serabyn, E. and Norris, B. and Tuthill, P. and Schworer, G. and Stewart, P. and Close, L. and Huby, E. and Perrin, G. and Lacour, S. and Gauchet, L. and Vievard, S. and Murakami, N. and Oshiyama, F. and Baba, N. and Matsuo, T. and Nishikawa, J. and Tamura, M. and Lai, O. and Marchis, F. and Duchene, G. and Kotani, T. and Woillez, J.},
   year={2015}, pages={890–910} }

@article{Wizinowich_Le_Mignant_Bouchez_Campbell_Chin_Contos_van_Dam_Hartman_Johansson_Lafon_et, title={The W. M. Keck Observatory Laser Guide Star Adaptive Optics System: Overview}, volume={118}, DOI={10.1086/499290}, abstractNote={The Keck Observatory began science observations with a laser guide star adaptive optics system, the first such system on an 8–10 m class telescope, in late 2004. This new capability greatly extends the scientific potential of the Keck II Telescope, allowing near–diffraction‐limited observations in the near‐infrared using natural guide stars as faint as 19th magnitude. This paper describes the conceptual approach and technical implementation followed for this system, including lessons learned, and provides an overview of the early science capabilities.}, number={840}, journal={Publications of the Astronomical Society of the Pacific}, publisher={The University of Chicago Press}, author={Wizinowich, Peter L. and Le Mignant, David and Bouchez, Antonin H. and Campbell, Randy D. and Chin, Jason C. Y. and Contos, Adam R. and van Dam, Marcos A. and Hartman, Scott K. and Johansson, Erik M. and Lafon, Robert E. and Lewis, Hilton and Stomski, Paul J. and Summers, Douglas M. and Brown, Curtis G. and Danforth, Pamela M. and Max, Claire E. and Pennington, Deanna M.}, year={2006}, month=feb, pages={297} }

@article{Beuzit_paper, title={SPHERE: the exoplanet imager for the Very Large Telescope}, volume={631}, DOI={10.1051/0004-6361/201935251}, journal={A\&A}, author={Beuzit, J.-L. and Vigan, A. and Mouillet, D. and Dohlen, K. and Gratton, R. and Boccaletti, A. and Sauvage, J.-F. and Schmid, H. M. and Langlois, M. and Petit, C. and Baruffolo, A. and Feldt, M. and Milli, J. and Wahhaj, Z. and Abe, L. and Anselmi, U. and Antichi, J. and Barette, R. and Baudrand, J. and Baudoz, P. and Bazzon, A. and Bernardi, P. and Blanchard, P. and Brast, R. and Bruno, P. and Buey, T. and Carbillet, M. and Carle, M. and Cascone, E. and Chapron, F. and Charton, J. and Chauvin, G. and Claudi, R. and Costille, A. and De Caprio, V. and de Boer, J. and Delboulbé, A. and Desidera, S. and Dominik, C. and Downing, M. and Dupuis, O. and Fabron, C. and Fantinel, D. and Farisato, G. and Feautrier, P. and Fedrigo, E. and Fusco, T. and Gigan, P. and Ginski, C. and Girard, J. and Giro, E. and Gisler, D. and Gluck, L. and Gry, C. and Henning, T. and Hubin, N. and Hugot, E. and Incorvaia, S. and Jaquet, M. and Kasper, M. and Lagadec, E. and Lagrange, A.-M. and Le Coroller, H. and Le Mignant, D. and Le Ruyet, B. and Lessio, G. and Lizon, J.-L. and Llored, M. and Lundin, L. and Madec, F. and Magnard, Y. and Marteaud, M. and Martinez, P. and Maurel, D. and Ménard, F. and Mesa, D. and Möller-Nilsson, O. and Moulin, T. and Moutou, C. and Origné, A. and Parisot, J. and Pavlov, A. and Perret, D. and Pragt, J. and Puget, P. and Rabou, P. and Ramos, J. and Reess, J.-M. and Rigal, F. and Rochat, S. and Roelfsema, R. and Rousset, G. and Roux, A. and Saisse, M. and Salasnich, B. and Santambrogio, E. and Scuderi, S. and Segransan, D. and Sevin, A. and Siebenmorgen, R. and Soenke, C. and Stadler, E. and Suarez, M. and Tiphène, D. and Turatto, M. and Udry, S. and Vakili, F. and Waters, L. B. F. M. and Weber, L. and Wildi, F. and Zins, G. and Zurlo, A.}, year={2019}, pages={A155} }

@inproceedings{REVOLT, title={REVOLT: an on-sky adaptive optics technology research platform on a 1.2m telescope}, volume={13097}, url={https://doi.org/10.1117/12.3019272}, DOI={10.1117/12.3019272}, note={Backup Publisher: International Society for Optics and Photonics}, booktitle={Adaptive Optics Systems IX}, publisher={SPIE}, author={Jackson, Kate and Conod, Uriel and Kooten, Maaike A. van and Véran, Jean-Pierre and Dunn, Jennifer and Lardière, Olivier and Kumar, Tarun and Chapin, Edward and Taylor, Jacob}, editor={Jackson, Kathryn J. and Schmidt, Dirk and Vernet, Elise}, year={2024}, pages={1309705} }

@article{Neichel_2014,
   title={Gemini multiconjugate adaptive optics system review – II. Commissioning, operation and overall performance},
   volume={440},
   ISSN={1365-2966},
   url={http://dx.doi.org/10.1093/mnras/stu403},
   DOI={10.1093/mnras/stu403},
   number={2},
   journal={Monthly Notices of the Royal Astronomical Society},
   publisher={Oxford University Press (OUP)},
   author={Neichel, Benoit and Rigaut, François and Vidal, Fabrice and van Dam, Marcos A. and Garrel, Vincent and Carrasco, Eleazar Rodrigo and Pessev, Peter and Winge, Claudia and Boccas, Maxime and d’Orgeville, Céline and Arriagada, Gustavo and Serio, Andrew and Fesquet, Vincent and Rambold, William N. and Lührs, Javier and Moreno, Cristian and Gausachs, Gaston and Galvez, Ramon L. and Montes, Vanessa and Vucina, Tomislav B. and Marin, Eduardo and Urrutia, Cristian and Lopez, Ariel and Diggs, Sarah J. and Marchant, Claudio and Ebbers, Angelic W. and Trujillo, Chadwick and Bec, Matthieu and Trancho, Gelys and McGregor, Peter and Young, Peter J. and Colazo, Felipe and Edwards, Michelle L.},
   year={2014},
   month=Apr, pages={1002–1019} }

@ARTICLE{2026arXiv260708146H,
       author = {{Haffert}, S.~Y. and {Liberman}, J. and {Males}, J.~R. and {Close}, L.~M. and {Foster}, W.~B. and {Van Gorkom}, K. and {Guyon}, O. and {Hedglen}, A.~D. and {Johnson}, P.~T. and {Kautz}, M.~Y. and {Kueny}, J.~K. and {Li}, J. and {Long}, J.~D. and {Lumbres}, J. and {Mars}, M. and {McEwen}, E.~A. and {McLeod}, A. and {Schatz}, L. and {Tonucci}, E. and {Twitchell}, K.},
        title = "{On-sky dark hole diggin' with implicit Electric Field Conjugation on MagAO-X}",
      journal = {arXiv e-prints},
      volume = {0},
         year = 2026,
        month = jul,
          eid = {arXiv:2607.08146},
        pages = {arXiv:2607.08146},
          doi = {10.48550/arXiv.2607.08146},
archivePrefix = {arXiv},
       eprint = {2607.08146},
 primaryClass = {astro-ph.IM},
       adsurl = {https://ui.adsabs.harvard.edu/abs/2026arXiv260708146H}
}

@inproceedings{Herriot_Morris_Anthony_Derdall_Duncan_Dunn_Ebbers_Fletcher_Hardy_Leckie_et, title={Progress on Altair: the Gemini North adaptive optics system}, volume={4007}, url={https://doi.org/10.1117/12.390288}, DOI={10.1117/12.390288}, note={Backup Publisher: International Society for Optics and Photonics}, booktitle={Adaptive Optical Systems Technology}, publisher={SPIE}, author={Herriot, Glen and Morris, Simon and Anthony, Andre and Derdall, Dennis and Duncan, Dave and Dunn, Jennifer and Ebbers, Angelic W. and Fletcher, J. Murray and Hardy, Tim and Leckie, Brian and Mirza, A. and Morbey, Christopher L. and Pfleger, M. and Roberts, Scott C. and Shott, Philip and Smith, M. and Saddlemyer, Leslie K. and Sebesta, Jerry and Szeto, Kei and Wooff, Robert and Windels, W. and Veran, Jean-Pierre}, editor={Wizinowich, Peter L.}, year={2000}, pages={115–125} }
\bibliographystyle{spiebib} 

\end{document}